\documentclass[
  aps,
  prb,
  reprint,
  superscriptaddress,
  floatfix,
  longbibliography
]{revtex4-2}

\usepackage[utf8]{inputenc}
\usepackage[T1]{fontenc}
\usepackage[english]{babel}

\usepackage{amsmath}
\usepackage{amssymb}
\usepackage{bm}

\usepackage{graphicx}
\usepackage{booktabs}

\graphicspath{{figuras/}}

\usepackage{microtype}

\usepackage[
  colorlinks=false,
  pdfborder={0 0 1},
  linkbordercolor={1 0 0},
  citebordercolor={0 1 0},
  urlbordercolor={0 1 1}
]{hyperref}

\hypersetup{
  pdftitle={Título del artículo},
  pdfauthor={Nombre del autor}
}

\begin{document}


\title{Configuration-Independent Low-Energy Gaps in Graphene Nanoribbons: A
		$\delta$-Scatterer Transfer-Matrix Approach}

\author{C.I. Avila-Joachin}
\thanks{Contact author:
  \href{mailto:autor@institucion.mx}{al2153070585@azc.uam.mx}}
\affiliation{%
Departamento de F\'isica, Universidad Aut\'onoma Metropolitana, Iztapalapa, Av. San Rafael Atlixco 186, Col. Vicentina, 09340, Ciudad de M\'exico, M\'exico
} 
\author{J.C. Sandoval-Santana}
\thanks{Contact author:
  \href{mailto:autor@institucion.mx}{jcss@azc.uam.mx}}
\author{Pedro Pereyra}
\thanks{Contact author:
  \href{mailto:autor@institucion.mx}{ppereyra@azc.uam.mx}}

\affiliation{%
\'Area de F\'isica Te\'orica y Materia Condensada, Universidad Aut\'onoma Metropolitana, Azcapotzalco, Av. San Pablo Xalpa 180, 02200, Ciudad de M\'exico, M\'exico
}
%

\date{\today}


\begin{abstract}
Transport measurements on lithographically patterned graphene nanoribbons suggest that the gap depends primarily on ribbon width rather than on crystallographic edge orientation, in contrast to predictions of ideal tight-binding models. To address this, we develop a transfer-matrix approach, based on the theory of finite periodic systems, in which each carbon atom of the honeycomb lattice is represented by an individual $\delta$-scatterer with interaction strength $\gamma^{}_{\delta}$ and the nanoribbon is treated as a confined multichannel finite periodic system. We derive the unit-cell transfer matrix for both arm-chair and zig-zag edge configurations and obtain the transmission and channel-coupling coefficients in closed form. The transmission spectra display the resonant band structure characteristic of finite periodic systems, with the number of resonances per allowed band increasing with the number of unit cells. For $\gamma_{\delta}=0$ the ribbons are fully transparent; for $\gamma_{\delta}>0$ a low-energy transport gap opens and grows linearly with the scatterer strength, $\Delta=0.364265\,\gamma_{\delta}$, identically for arm-chair and zig-zag edges. These results are consistent with the reported absence of a systematic dependence on orientation, while the analytical results show that interchannel coupling vanishes exactly for arm-chair ribbons and decreases with increasing width for zig-zag ribbons, revealing the role of edge geometry. Together, these findings indicate that atomic-scale scattering, represented by the $\delta$-potential formalism, provides a common orientation-independent contribution to the low-energy transport gap in the pristine ribbons considered here.
\end{abstract}

\maketitle

\section{Introduction}
\label{sec:introduccion}

Soon after the experimental isolation of monolayer graphene \cite{novoselov2004}, it was recognized that much of its low-energy electronic structure had already been anticipated by Wallace decades earlier \cite{wallace1947}, in his work on graphite. In that work, the interlayer coupling was initially neglected; consequently, the obtained band structure corresponds to an isolated carbon layer, now identified with graphene. Based on a tight-binding approximation and Bloch-type wave functions, Wallace showed that near the corners of the hexagonal Brillouin zone, the energy varies linearly with the wave-vector displacement from the corner point, and that the energy gap between the two bands vanishes as a consequence of the lattice symmetry. This linear dispersion near the corners, now identified with the Dirac cone and Dirac points, was later investigated experimentally through the observation of massless Dirac-like quasiparticles, evidenced by the relation between cyclotron mass and carrier concentration \cite{novoselov2005}, and through the anomalous, half-integer quantum Hall effect \cite{zhang2005}. These and other striking electronic properties of graphene have been reviewed in \cite{castroneto2009}.

In this context, Han et al.\ \cite{han2007} investigated electronic transport in lithographically patterned graphene nanoribbons (GNRs) of varying widths and crystallographic orientations. Their main experimental finding is that the gap scales inversely with the ribbon width, following the empirical relation
$E_{gap} = \alpha/(W-W^*)$, where $\alpha = 0.2$ eV$\cdot$nm and $W^* = 16$ nm. These results were obtained from a linear fit to the data, showing that lateral confinement plays a central role in opening a gap in GNRs. However, Han et al.\ also reported that no systematic crystallographic-orientation dependence was observed in the gap, despite theoretical predictions dating back to the tight-binding analysis of edge states by Nakada et al.\ \cite{nakada1996} and later refined by Son et al.\ \cite{son2006}, which suggested that zig-zag and arm-chair ribbons should exhibit qualitatively different electronic behavior. Furthermore, Han et al.\ pointed out that the interplay between edge structure and chemical termination remains an open question. The experimentally observed gap may therefore not be fully captured by simple lateral-confinement models. It has been suggested that additional intrinsic scattering mechanisms may also contribute, a possibility later supported by direct experimental evidence of electron--electron interactions contributing to gap opening in zig-zag graphene nanoribbons \cite{li2014}.

In a related experiment, Stampfer et al.\ \cite{stampfer2009} performed detailed transport measurements on an etched graphene nanoribbon of width $W \approx 45$ nm, finding a transport gap of $E_g \approx 8$ meV. They explicitly noted that standard theoretical models, including zone-folding approximations, $\pi$-orbital tight-binding, and first-principles calculations, could not explain the experimentally observed energy gaps for ribbon widths beyond 20 nm. This discrepancy points to the possibility that, beyond local interaction features, additional physical mechanisms associated with the global periodic structure of the system may contribute to the transport gap. It is well known that phase coherence in periodic systems can give rise to band structures. This motivates analyzing graphene nanoribbons from the perspective of finite periodic systems.

Our purpose here is to study electronic transport across laterally confined graphene nanoribbons, taking into account both their lateral confinement and the periodic arrangement of the carbon atoms. Since the honeycomb lattice contains two fundamental crystallographic sequences, the arm-chair and zig-zag sequences, we consider a model and theoretical treatment in which the global periodic structure, the crystallographic orientation, the lateral confinement, and the strength of the particle-array interactions are explicitly included. This allows the transport band structure and the opening of a gap in electronic transport across a graphene nanoribbon to be calculated within a finite periodic-system approach.

The formalism proposed in this work, based on the theory of finite periodic systems (TFPS) \cite{pereyra2002,pereyra1998,pereyra2000}, offers a complementary description of the transport gap. Transfer-matrix methods have been applied to graphene waveguides, superlattices, and nanoribbons~\cite{Li2009,Wang2014,Wakabayashi2010,Doost2016}. Unlike conventional approaches based on tight-binding or effective Dirac Hamiltonians, here we emphasize the role of the global periodic structure, the crystallographic orientation, the lateral confinement, and the strength of the particle-array interactions. For this purpose, we define crystallographic transport-unit-cell stripes of carbon atoms whose transverse dimension is the same as the transverse width of the GNR and whose length is similar to that of a hexagon of the honeycomb array (see Fig.~\ref{ArmGraphene}). At the atomic positions, an effective $\delta$-scatterer potential of intrinsic strength $\gamma^{}_{\delta}$ is assumed. Together with the transport unit cells, we introduce the multichannel finite periodic-system approach and calculate, analytically and numerically, the transmission coefficients as functions of the longitudinal particle energy. We restrict the analysis to energies with two propagating modes, also known as open channels, defined in the leads connected to the GNR.

Our results show that, while the transport band structure depends on the nanoribbon orientation with respect to the crystallographic axes, the low-energy transport gap obtained in the present model is independent of the crystallographic orientation of the ribbon. The gap is generated by the periodic $\delta$-scatterer potential within the laterally confined ribbon model, rather than by lateral confinement alone. This result is consistent with the experimental observation of Han et al.\ that no systematic dependence on crystallographic direction is found.

The paper is organized as follows. Section \ref{sec:model} introduces the transfer-matrix approach for a graphene nanoribbon, defining the confining potential, the array of $\delta$-scatterer centers that encode the honeycomb lattice, and the atomic positions for the zig-zag and arm-chair edge configurations. We define the transport unit cells for zig-zag and arm-chair configurations, and derive the unit-cell transfer matrices and their relation to the scattering amplitudes within the theory of finite periodic systems. Section \ref{sec:results} presents the numerical results: the resonant transmission structure as a function of the number of unit cells, the opening of the band gap as a function of the scatterer strength $\gamma^{}_\delta$, and the scaling relations extracted for the arm-chair and zig-zag sequences at the GNR edges. Section \ref{sec:conclusions} summarizes our findings.

\section{The TM approach for a graphene nanoribbon}\label{sec:model}

\begin{figure}[ht]
  \centering
\includegraphics[width=\linewidth]{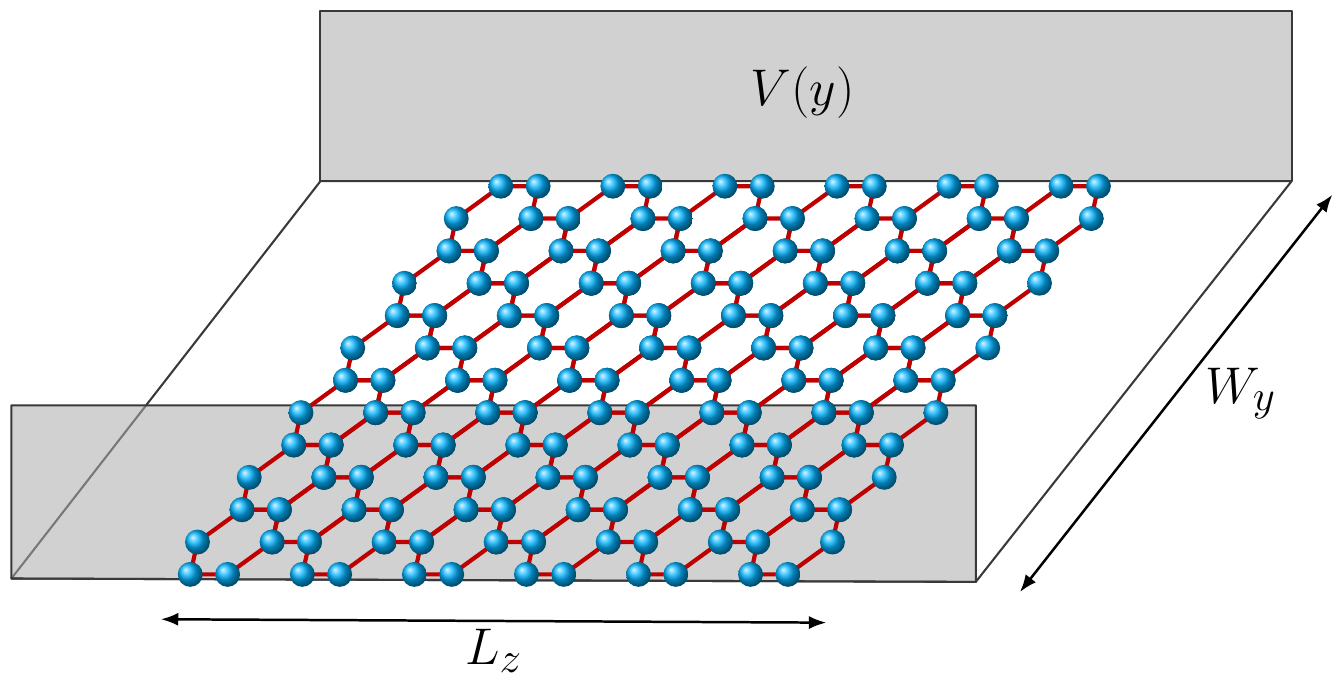}
  \caption{A 2D graphene nanoribbon confined by an infinite square-well potential in the $y$-direction and extended along the $z$-direction.}\label{ConfinedGraphene}
\end{figure}

Differently from conventional formulations based on tight-binding models or effective Dirac Hamiltonians to describe the low-energy states near the $K$ and $K'$ points, here we adopt a real-space multichannel scattering description in which the periodicity of the honeycomb lattice is encoded through the positions of atomic $\delta$-scatterers rather than through the band Hamiltonian itself. These scatterers are placed at the carbon-atom sites of the hexagonal array and are characterized by an effective interaction strength $\gamma^{}_\delta$. We consider injected charge carriers with energies around the Fermi level. At low energies measured with respect to the Fermi level, the particles experience essentially a constant effective potential, while the honeycomb periodicity is represented by effective repulsive (positive $\gamma^{}_\delta$) or attractive (negative $\gamma^{}_\delta$) interactions at the atomic sites. Within this physical picture, we calculate the transport properties across a system of length $L_z$ and width $W_y$ as functions of the particle energy, the interaction strength, the GNR width, the number of unit cells $n$, and the crystallographic orientation.

We consider the array of scattering centers inside a confining potential $V_c(y)$ given by
\begin{equation}
	V_c(y) =
	\begin{cases}
		\infty, & y \leq 0 \\
		0, & 0 < y < W_y \\
		\infty, & y \geq W_y
	\end{cases} \, ,
\end{equation}
which extends to the left and right of the sample as a waveguide (see Fig.~\ref{ConfinedGraphene}) and confines the incoming and outgoing electrons as well as the particles inside the sample.

\begin{figure}[ht]
	\begin{center}
		\includegraphics[angle=0,width=0.48\textwidth]{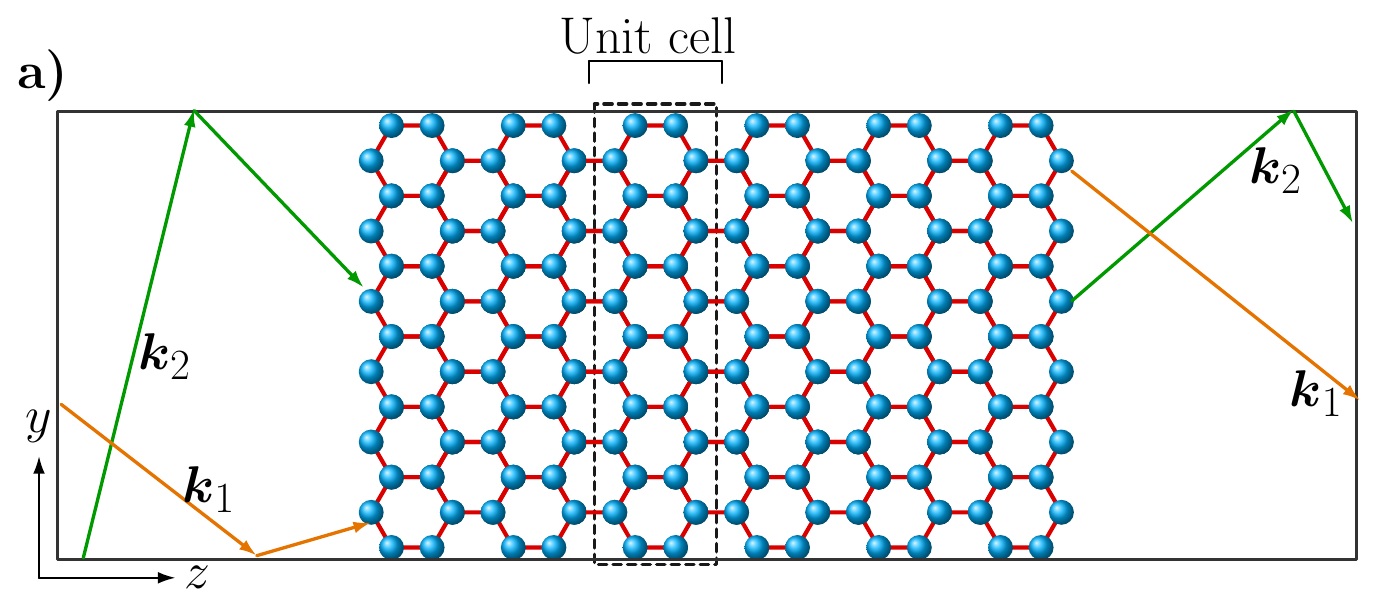}
		\includegraphics[angle=0,width=0.48\textwidth]{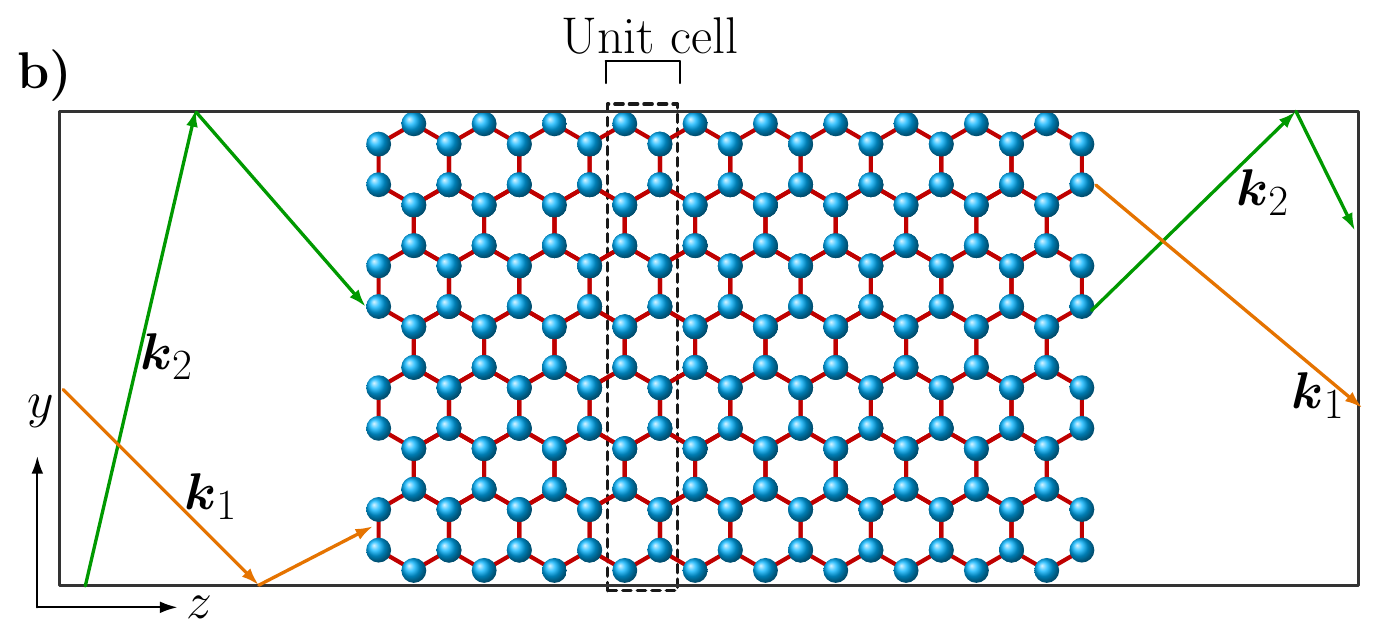}
		\caption{Graphene nanoribbons in (a) arm-chair and (b) zig-zag configurations. The dashed lines delimit the unit cells used to construct the transfer matrix: the arm-chair unit cell contains four lines of atoms {\it in two zig-zag sequences} along the transverse direction, while the zig-zag unit cell contains two lines of atoms {\it in two arm-chair sequences}. The leads at the left and right of the sample impose quantization of the transverse motion of the incoming, reflected, and transmitted particles.}\label{ArmGraphene}
	\end{center}
\end{figure}

As mentioned before, the key ingredient in the transfer-matrix approach is the proper definition of the transport unit cell. This cell is different from the crystallographic unit cell of graphene. While the graphene unit cell is defined by the lattice periodicity, in the transport description along the $z$-axis of a two-dimensional system, the transport unit cell is defined as the minimal sample stripe of width $W_y$ and length $l_c$. Repetition of this unit cell $n$ times along the transport direction forms a sample of length $L_z=n l_c$ and width $W_y$. In this approach, the accurate calculation of the unit-cell transfer matrix determines the accuracy of the total transfer matrix and of the ensuing results.
For arm-chair and zig-zag configurations, the transport unit cell is composed of four and two full transverse lines of atoms, respectively, as shown in Figs.~\ref{ArmGraphene} a) and b), where the unit cells are enclosed by the dashed frames. Notice that the arm-chair transport-unit cell contains four lines of atoms in two zig-zag sequences, while the zig-zag unit cell contains two lines of atoms in two arm-chair sequences along the $y$-axis. The transverse atomic positions $y_{l,\mu}$ within each transport unit cell depend on the edge configuration. For the arm-chair configuration, atoms in odd lines $y_{1,\mu}$ and even lines $y_{2,\mu}$ are given by

\begin{equation}\label{posAC}
	y_{1,\mu} =(2\mu+1)\,a\frac{\sqrt{3}}{2} \hspace{.3 cm} \text{and} \hspace{.3 cm} y_{2,\mu} = \mu\,a\sqrt{3} \, .
\end{equation}
Here, the indices $l$ and $\mu$ label the atomic line and the atom within that line, respectively. For the zig-zag configuration, the positions are given by
\begin{equation}
\label{posZZ}
\begin{aligned}
y_{1,\mu} &=
\begin{cases}
\dfrac{3\mu+1}{2}\,a, & \text{if } \mu \text{ even}, \\[6pt]
\dfrac{3\mu}{2}\,a,   & \text{if } \mu \text{ odd},
\end{cases}
\\[12pt]
y_{2,\mu} &=
\begin{cases}
\dfrac{3\mu}{2}\,a,   & \text{if } \mu \text{ even}, \\[6pt]
\dfrac{3\mu+1}{2}\,a, & \text{if } \mu \text{ odd}.
\end{cases}
\end{aligned}
\end{equation}
and encode the alternating bond structure characteristic of the zig-zag edge.

To obtain the transfer matrix of the transport unit cell, we solve the Schrödinger equation within this cell, whose Hamiltonian is
\begin{equation}
	H=-\frac{\hbar^2}{2m_e^*}\left( \frac{\partial^2}{\partial y^2}+ \frac{\partial^2}{\partial z^2}\right)+V_c(y)+V_{p}(y,z) \, ,
\end{equation}
where $m_e^*$ is the electron effective mass. The term $V_c(y)$ represents the infinite square-well potential that confines the electron along the $y$-direction. The potential $V_p(y,z)$ accounts for the interaction between the electron and the array of scattering centers of the graphene lattice, arranged according to the crystallographic symmetry, and is modeled as
\begin{equation}
	V_{p}(y,z)= \sum_{l,\mu}^{N_l,N_\mu} \gamma^{}_\delta \delta\left(y-y_{l,\mu}\right) \delta \left(z-z_{l}\right) \, ,
\end{equation}
where $\gamma^{}_\delta$, as mentioned before, is the strength of the interaction.

To describe the electronic transport of particles through the scattering configurations shown in Figs.~\ref{ArmGraphene}, it is convenient to define first the propagating modes in the leads, i.e., at the left and right sides of the sample. For a given incoming particle energy $E$, relative to the Fermi energy, the propagating modes are defined as those whose wave numbers satisfy the relation
\begin{equation}\label{Graf3}
	{k}_{li}^2 = \frac{2m_e^*E}{\hbar^2}-{k}_{Ti}^2 \geq 0.
\end{equation}
where $k_{Ti}=n_i\pi/W_y$ and $k_{li}$ are the transverse and longitudinal wave numbers, respectively, with $n_i$ integer numbers. Following Bagwell \cite{Bagwell}, the wave function $\psi(y,z)$ can be expressed as the linear combination

\begin{equation}\label{Graf5}
	\psi\left(y,z\right) =\sum _{i}\varphi _{i}(y) \phi_{i}(z).
\end{equation}
where
\begin{equation}\label{Graf4}
	\varphi_{i}(y)=\sqrt{\frac{2}{W_y}}\sin{\frac{n_{i}\pi}{W_y}y},
\end{equation}
are the eigenfunctions in the leads.
If we introduce this function in the Schr\"{o}dinger equation, multiply by $\varphi^*_{i}(y)$ and integrate, we obtain the system of coupled equations
\begin{equation}
\label{eq:coupled}
\begin{aligned}
\frac{d^{2}\phi_i(z)}{dz^{2}}
&+ \left(k^{2}-k_{Ti}^{2}\right)\phi_i(z)
\\
&= \sum_j K_{l,ij}(z)\,\phi_j(z).
\end{aligned}
\end{equation}
Here $i,j=1,2,\ldots$ and $l=1,\ldots,N_l$, with
$k=(2m_e^*E/\hbar^2)^{1/2}$ and
$k_{Ti}=(2m_e^*E_{Ti}/\hbar^2)^{1/2}$.
The coupling matrix elements are
\begin{equation}
\label{Graf7}
\begin{aligned}
K_{l,ij}(z)
&= \frac{2m_e^*}{\hbar^{2}}
   \int \varphi_i^*(y)\,
   V_P(y,z)\,\varphi_j(y)\,\mathrm{d}y
\\[4pt]
&= \Gamma_{l,ij}\,\delta(z-z_l),
\end{aligned}
\end{equation}
where 
\begin{equation}\label{eqGammaijl}
	\Gamma_{l,ij}=\frac{4m^*_{e}}{\hbar^2 W_{y}} \gamma^{}_{\delta} \sum_{\mu}\sin{\left(\frac{n_{i}\pi}{W_{y}}y _{l,\mu}\right)} \sin{\left(\frac{n_{j}\pi}{W_{y}}y_{l,\mu}\right)}\, .
\end{equation}

For a given line of $\delta$ scatterers located at $z=z_l$, Eq.~(\ref{eq:coupled}) can be written as
\begin{equation}\label{eq:coupled_delta}
	\frac{d^2\varphi_i(z)}{dz^2}
	+k_i^2\varphi_i(z)
	=
	\sum_j
	\Gamma_{l,ij}\delta(z-z_l)\varphi_j(z),
\end{equation}
where $k_i^2=k^2-k_{T_i}^2$.
Integrating Eq.~(\ref{eq:coupled_delta}) over an infinitesimal interval
around $z_l$ gives
\begin{equation}\label{DerivContinuity}
	\varphi_i'(z_l^+)-\varphi_i'(z_l^-)
	=
	\sum_j \Gamma_{l,ij}\varphi_j(z_l),
\end{equation}
Here $z_l^{\pm}= \lim_{\epsilon \rightarrow 0} \,(z_l{\pm}\epsilon)$. (\ref{DerivContinuity}) defines the discontinuity of the wave functions derivatives, while the wave functions continuity is given by
\begin{equation}
	\varphi_i(z_l^+)=\varphi_i(z_l^-).
	\label{eq:continuity}
\end{equation}

The longitudinal wave functions represent in general the superposition of right- and
left-moving waves. Thus, at the left and right sides of $z_l$, we have
\begin{equation}
\label{eq:LRwave}
\begin{aligned}
\varphi_i(z<z_l)
&= A_i^- e^{ik_i(z-z_l)}
 + B_i^- e^{-ik_i(z-z_l)}, \\[6pt]
\varphi_i(z>z_l)
&= A_i^+ e^{ik_i(z-z_l)}
 + B_i^+ e^{-ik_i(z-z_l)}.
\end{aligned}
\end{equation}

Evaluating Eq. (\ref{eq:LRwave}) at $z=z_l$, the continuity condition
(\ref{eq:continuity}) becomes
\begin{equation}
	A_i^+ + B_i^+
	=
	A_i^- + B_i^- .
	\label{eq:amplitude_continuity}
\end{equation}
Similarly, the derivatives continuity condition in  Eq.~(\ref{DerivContinuity}) yields
\begin{equation}
	ik_i
	\left(
	A_i^+-B_i^+-A_i^-+B_i^-
	\right)
	=
	\sum_j
	\Gamma_{l,ij}
	\left(
	A_j^-+B_j^-
	\right).
\end{equation}
These two  boundary conditions can be combined  into a compact matrix form
\begin{equation}
	\begin{pmatrix}
		A_1^+\\
		A_2^+\\
		B_1^+\\
		B_2^+
	\end{pmatrix}
	=
	M_{\delta,l}
	\begin{pmatrix}
		A_1^-\\
		A_2^-\\
		B_1^-\\
		B_2^-
	\end{pmatrix},
	\label{eq:delta_transfer_relation}
\end{equation}
where the scattering transfer matrix  is
\begin{equation}
	M_{\delta,l}
	=
	\begin{pmatrix}
		1-i\dfrac{\Gamma_{l,11}}{2k_1}
		&
		-i\dfrac{\Gamma_{l,12}}{2k_1}
		&
		-i\dfrac{\Gamma_{l,11}}{2k_1}
		&
		-i\dfrac{\Gamma_{l,12}}{2k_1}
		\\[2mm]
		-i\dfrac{\Gamma_{l,21}}{2k_2}
		&
		1-i\dfrac{\Gamma_{l,22}}{2k_2}
		&
		-i\dfrac{\Gamma_{l,21}}{2k_2}
		&
		-i\dfrac{\Gamma_{l,22}}{2k_2}
		\\[2mm]
		i\dfrac{\Gamma_{l,11}}{2k_1}
		&
		i\dfrac{\Gamma_{l,12}}{2k_1}
		&
		1+i\dfrac{\Gamma_{l,11}}{2k_1}
		&
		i\dfrac{\Gamma_{l,12}}{2k_1}
		\\[2mm]
		i\dfrac{\Gamma_{l,21}}{2k_2}
		&
		i\dfrac{\Gamma_{l,22}}{2k_2}
		&
		i\dfrac{\Gamma_{l,21}}{2k_2}
		&
		1+i\dfrac{\Gamma_{l,22}}{2k_2}
	\end{pmatrix}.
	\label{eq:Mdelta}
\end{equation}
The $4\times4$ dimension reflects the two propagating modes considered in the system, each with right- and left-propagating amplitudes. The coefficients $\Gamma_{l,ij}$, with $i,j=1,2$, describe the coupling between the propagating modes at the scatterer line, with characteristic wave vectors $k_1$ and $k_2$. To propagate the state vector from $z_l$ to $z_{l+1}$, one also needs the free-propagation transfer matrix.
\begin{eqnarray}
	m_{1}&=&\begin{pmatrix}
		e^{ik_{1} d/2}& 0 & 0 & 0 \\
		0 & e^{ik_{2} d/2} & 0 & 0 \\
		0 & 0 & e^{-ik_{1} d/2} & 0\\
		0 & 0 & 0 & e^{-ik_{2}d/2}\\
	\end{pmatrix}
\end{eqnarray}
where $d=a,a\sqrt{3}$ for  arm-chair and zig-zag configurations, respectively. Here $a$ is the interatomic distance in graphene. The unit-cell transfer matrix for arm-chair and zig-zag configurations are given by
\begin{equation}\label{eq:MAC}
	M_{AC}=m_1M_{\delta,1}m_1M_{\delta,2}m_1m_1M_{\delta,2}m_1M_{\delta,1}m_1 \,
\end{equation}
and
\begin{equation}
	M_{ZZ}=m_1M_{\delta,1}m_1m_1M_{\delta,2}m_1 \, ,
\end{equation}
respectively. To obtain the transfer matrix of the whole system and the transmission coefficients, we  use the closed formulas of the theory of finite periodic systems \cite{pereyra2002,pereyra1998}. In this theory the transfer matrix of the $n$ unit-cells system is given by
\begin{equation}
	M_n=M^n=\begin{pmatrix} \alpha & \beta\\ \gamma & \delta\end{pmatrix}^n = \begin{pmatrix} \alpha_n & \beta_n\\ \gamma_n & \delta_n\end{pmatrix},
\end{equation}
where
\begin{equation}
	\alpha_n=p_n-\gamma^{-1}\delta \gamma p_{n-1} \hspace{0.2in}{\rm and}\hspace{0.2in}\beta_n=\beta p_{n-1}
\end{equation}
with the matrix polynomials $p_n$ determined by the matrix-recurrence relation
\begin{equation}\label{Graph14}
	p_n-(\beta^{-1}\alpha \beta+\delta) p_{n-1}+ (\delta \beta^{-1}\alpha \beta-\gamma \beta)p_{n-2}=0
\end{equation}
whose solutions are given in Refs. \cite{pereyra2002,pereyra1998}. These polynomials reduce to the Chebyshev polynomials in the scalar, one propagating mode case. The relations between the transfer matrix blocks and the scattering amplitudes is direct, and given by
\begin{equation}\label{Graph15}
	t=(\alpha^{\dagger})^{-1}\hspace{0.2in}{\rm and}\hspace{0.2in} r=(\alpha^{*})^{-1}\beta^*.
\end{equation}
In this approach for a two open channels description, it is easy to calculate the transmission coefficients. Since the transmission amplitude is
\begin{equation}
	t = \frac{1}{\det \alpha_n^T}
	\begin{pmatrix}
		\alpha_{n,22} & -\alpha_{n,21} \\
		-\alpha_{n,12} & \alpha_{n,11}
	\end{pmatrix}^*,
\end{equation}
the transmission coefficient is
\vspace{.5cm}

\begin{equation}
\label{eq:transmission-matrix}
T = tt^\dagger
= \frac{1}{|\det \alpha_n|^2}
\begin{pmatrix}
\mathcal{A}_n & \mathcal{B}_n \\
\mathcal{B}_n^* & \mathcal{C}_n
\end{pmatrix},
\end{equation}
where
\begin{equation}
\label{eq:transmission-elements}
\begin{aligned}
\mathcal{A}_n
&= |\alpha_{n,22}|^2 + |\alpha_{n,21}|^2, \\[4pt]
\mathcal{B}_n
&= -\alpha_{n,21}\alpha_{n,22}^*
   -\alpha_{n,11}\alpha_{n,12}^*, \\[4pt]
\mathcal{C}_n
&= |\alpha_{n,11}|^2 + |\alpha_{n,12}|^2.
\end{aligned}
\end{equation}

\section{Results}\label{sec:results}

In this section, we present the results of our transfer-matrix calculations for graphene nanoribbons with arm-chair and zig-zag edge configurations. For the zig-zag case, the width and length are $W_y = 3N_y a$ and $L_z = 2na$, respectively, while for the arm-chair case they are $W_y = \sqrt{3}N_y a$ and $L_z = 3na$. Our analysis proceeds in four stages. First, we determine the resonant transmission coefficients for arm-chair and zig-zag configurations for two values of the number of unit cells. Second, we analyze the effect of the sign of the interaction strength $\gamma^{}_\delta$ on the opening of the energy gap and on the transmission-band widths. We then examine how the gap opens as a function of the interaction strength $\gamma^{}_\delta$, and finally extract the scaling relations between the gap and the interaction strength for both edge configurations. Due to the periodic structure of the system, the coupling coefficients $\Gamma_{l,ij}$ can be calculated analytically. In Appendix A, we present the detailed derivation.

\begin{figure*}[ht]
	\begin{center}
		\includegraphics[angle=0,width=0.39\textwidth]{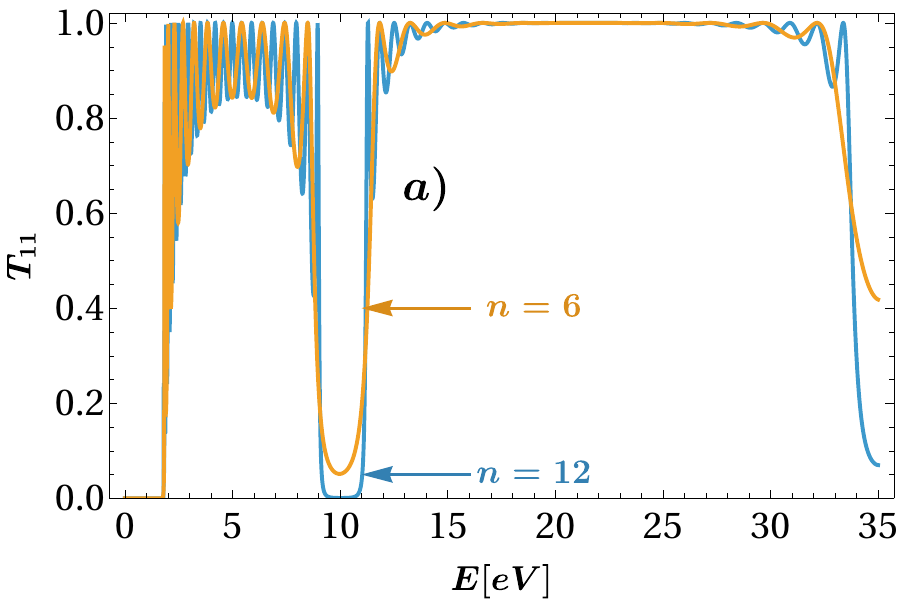}
		\includegraphics[angle=0,width=0.39\textwidth]{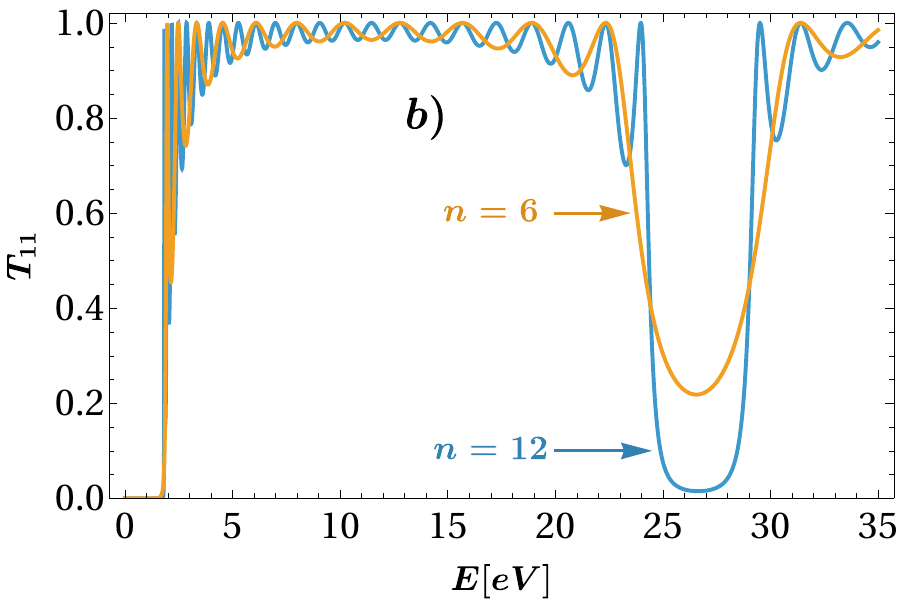}
		\caption{Transmission coefficient $T_{11}$ as a function of the energy $\text{E}$ for graphene nanoribbons with $n=6$ (orange curve) and $n=12$ (blue curve) unit cells, considering two propagating modes ($N=2$) for $\gamma^{}_\delta =5$. The a) and b) panels correspond to the arm-chair ($N_y = 253$) and zig-zag ($N_y = 146$) edge configurations, respectively. In both cases, $T_{11}$ denotes the probability that an electron injected in channel 1 is transmitted through the same channel.}\label{TACZZ}
	\end{center}
\end{figure*}

In Fig.~\ref{TACZZ} we show the transmission coefficient $T_{11}$ as a function of the energy $E$, for $\gamma^{}_{\delta} = 5$ and two propagating modes (${\cal N}=2$), in
graphene nanoribbons with $n = 6$ (orange curve) and $n = 12$ (blue curve) unit cells.
The nanoribbon lengths along the transport direction are $L_z \approx 1.7$~nm ($n=6$) and
$L_z \approx 3.4$~nm ($n=12$) for the zig-zag configuration,
and $L_z \approx 2.5$~nm ($n=6$) and
$L_z \approx 5$~nm ($n=12$) for the arm-chair configuration. The arm-chair and zig-zag
edge configurations correspond to transverse widths
$N_y = 253$ and $N_y = 146$, respectively, both corresponding to
$W_y \approx 62$~nm. Here, $T_{11}$ represents the probability that an electron
injected in channel~1 is transmitted through the same channel.

\begin{figure*}[ht]
	\begin{center}
		\includegraphics[angle=0,width=14.5cm]{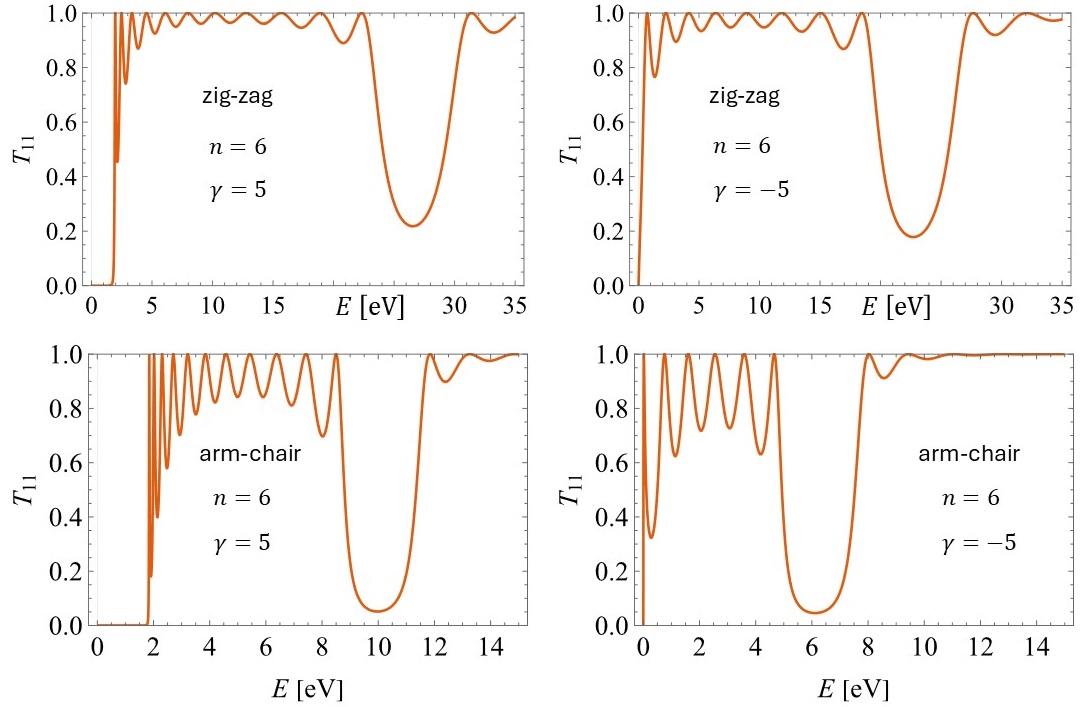}
		\caption{Transmission coefficient $T_{11}$ as a function of the energy $\text{E}$ for positive and negative $\gamma^{}_\delta$. In the upper panels for zig-zag and lower panels for arm-chair configurations. On the left column, the transmission coefficients shown also in Fig.~\ref{TACZZ} for $n=$6 and positive $\gamma^{}_\delta=5$. On the right column, the transmissions correspond to negative $\gamma^{}_\delta=-5$. While the gaps open for positive $\gamma^{}_\delta$, they close for negative $\gamma^{}_\delta$.}\label{TACZZGamma}
	\end{center}
\end{figure*}

Panels (a) and (b) correspond to arm-chair and zig-zag configurations, respectively. The transmission coefficient displays the resonant band structure characteristic of a finite periodic system, with the number of resonances per allowed band equal to $2n-1$, in agreement with the theory of finite periodic systems \cite{pereyra2002}. This relation follows from the fact that each transport unit cell contains two wells, so that a system composed of $n$ unit cells contains $2n$ wells and therefore $2n-1$ transmission resonances within each allowed band. As $n$ increases from 6 to 12, the number of intraband resonances increases accordingly, reflecting the increasing role of phase coherence in the finite periodic system.

The oscillatory structure of $T_{11}$ within the allowed bands results from the constructive and destructive interference of the phase-coherent propagating and scattered waves. The propagation phase accumulated across the finite periodic structure depends on the particle energy through the de Broglie wavelength. As in simpler resonant systems such as quantum wells and double barriers, transmission resonances occur when the accumulated phase satisfies the constructive-interference condition, corresponding to an effective propagation length that is an integer multiple of $\lambda_B/2$. Increasing the number of unit cells therefore produces additional transmission resonances within the allowed bands.

\begin{figure*}[ht]
	\begin{center}
		\includegraphics[angle=0,width=14.7cm]{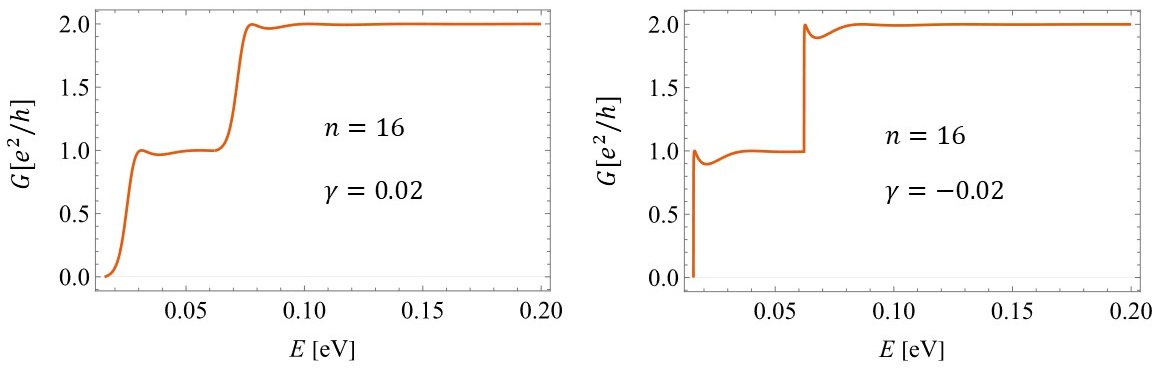}
		\caption{Conductance $G \propto {\rm Tr}tt^{\dagger}$ as function of the energy, in units of $e^2/h$ for $n=16$, $\gamma^{}_\delta=$0.02, in the left panel, and $\gamma^{}_\delta=-$0.02, in the right panel. }\label{Conductance}
	\end{center}
\end{figure*}

\begin{figure*}[ht]
	\begin{center}
		\includegraphics[angle=0,width=0.43\textwidth]{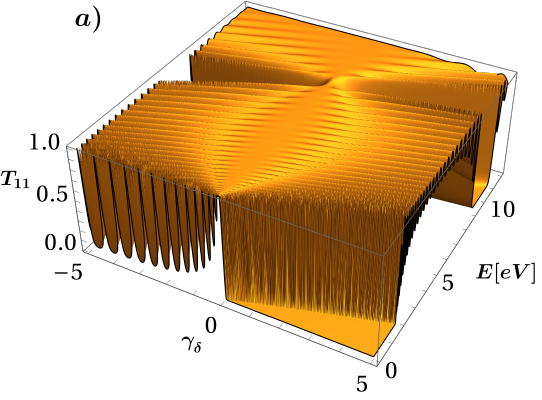}
		\includegraphics[angle=0,width=0.43\textwidth]{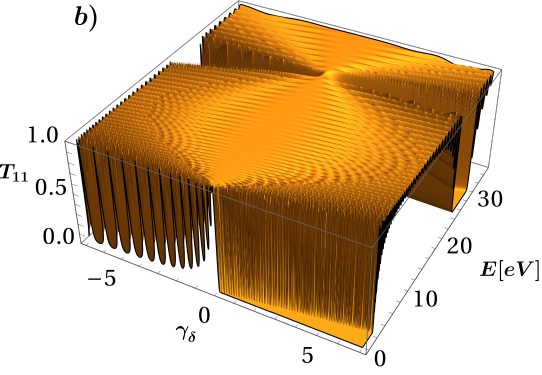}
		\caption{Transmission coefficient as a function of the coupling strength $\gamma^{}_\delta$ and energy for $n$ unit cells oriented along the transport direction: (a) arm-chair configuration with a ribbon width of $N_y = 253$ with $n=15$ and (b) zig-zag considering $N_y = 146$ with $n=22$. The band structure is sensitive to $\gamma^{}_\delta$. For $\gamma^{}_\delta = 0$, the periodic scattering potential is absent, the scatterer-induced band structure disappears, the gap vanishes, and the sample is fully transparent ($T_{11} = 1$). For $\gamma^{}_\delta > 0$, the gap increases linearly with $\gamma^{}_\delta$ (see lower corner). For $\gamma^{}_\delta < 0$, the first gap appears only at relatively large energy values for both configurations.
		}\label{TACZZ3D}
	\end{center}
\end{figure*}

For $E\lesssim 2$~eV, and for both arm-chair and zig-zag configurations, a band gap is observed, where the transmission coefficients are completely suppressed. While the band-gap widths are equal for arm-chair and zig-zag configurations, the allowed-band widths beyond this region, differ markedly. The arm-chair nanoribbon presents a forbidden-energy region at $E\approx 9$--$10$~eV, whereas for the zig-zag nanoribbon the corresponding forbidden region is shifted to considerably higher energies, $E\approx 25$--$29$~eV.  The pronounced suppression of the transmission coefficients is deeper and better defined when the number of unit cells increases. This is well known property in finite periodic systems.

Fig.~\ref{TACZZGamma} shows the transmission coefficient $T_{11}$ as a function of the
energy $\text{E}$ for positive and negative $\gamma^{}_\delta$. In this figure we consider zig-zag and arm-chair nanoribbons with the same geometry as in Fig.~\ref{TACZZ} for $n=$6.  In the upper panels the transmissions for zig-zag configuration are shown and in the lower panels for arm-chair nanoribbon. The graphs on the left are for $\gamma^{}_\delta=$5, and  those on the right panels are for $\gamma^{}_\delta=-$5. It is clear from these figures that gaps open when $\gamma^{}_\delta$ is positive, and they close when $\gamma^{}_\delta$ is negative. The values of \(\gamma_\delta\) used here are deliberately large, producing correspondingly wide gaps at relatively high energies. These values are therefore useful for illustrating the effect of the interaction strength, but they are not intended as a direct representation of the interaction strength in a realistic graphene nanoribbon.

To check whether the behavior seen in Fig. \ref{TACZZGamma} remains or not for much smaller interaction strengths, we plot in Fig. \ref{Conductance} the conductance $G \propto {\rm Tr}tt^{\dagger}$, in units of $e^2/h$ for $n=16$, $\gamma^{}_\delta=$0.02, in the left panel, and $\gamma^{}_\delta=-$0.02, in the right panel. These graphs show that for positive interaction strength the lower-energy channel shows a slow growth of the transmission coefficient, while for negative interaction strength there is no gap energy before the onset of the transmission coefficient, as for larger interaction strengths. For negative interaction strength, the allowed transport bands become thinner and shift to lower energies. The first allowed band is sharply cut at zero energy, and the number of resonances is reduced.

In Fig. \ref{TACZZ3D} a) and  b), we present the transmission coefficients $T_{11}(\gamma^{}_\delta,\text{E})$ as a function of the strength of the interaction and the energy for the arm-chair and zig-zag configurations, respectively. The arm-chair system consists of $n = 15$ unit cells with $N_y = 253$, while the zig-zag system comprises $n = 22$ unit cells with $N_y = 146$. These correspond approximately to dimensions of $L_z = 6\,\mathrm{nm}$ and $W_y = 62\,\mathrm{nm}$ for both configurations. Within the present calculation, the band gap vanishes at $\gamma^{}_\delta = 0$, which implies no interaction and complete transmission for these systems. We see that for an effective repulsive interaction, i.e. for $\gamma^{}_\delta > 0$, the gap opens up as a function of the strength $\gamma^{}_\delta$. The band gap observed in both configurations are consistent with theoretical predictions and experimental observations \cite{han2007,son2006,li2014,stampfer2009}. In this approach we have only one parameter, which is $\gamma^{}_\delta$. This parameter can be fixed with experimental measurements of the gap.

To determine the relationship between the band gap and $\gamma_\delta$, we summarize the main result of Appendix A. The coupling coefficients can be written as $\Gamma_{l,ij}=(4m_e^{*}\gamma_\delta/\hbar^2W_y)\Lambda_{l,ij}$, where $\Lambda_{l,ij}$ is the overlap between the transverse modes evaluated at the atomic positions of the scatterer line. For the arm-chair configuration the channels are exactly decoupled, $\Lambda_{l,ij}=0$ for $i\neq j$, whereas for the sufficiently wide zig-zag ribbons considered here the interchannel coupling becomes negligible. In both configurations, $\Lambda_{l,ii}\propto N_y$, and therefore the diagonal coupling is proportional to $\gamma_\delta$. Thus, for the ribbons considered here, the channels are effectively decoupled and $M_{\delta,l}$ and $m_1$ are block diagonal. Consequently, the transport unit cell matrices $M_{\mathrm{AC}}$ and $M_{\mathrm{ZZ}}$ have the same $2\times2$ block structure, allowing each channel to be treated independently.

\begin{figure}[ht]
  \centering
\includegraphics[width=\linewidth]{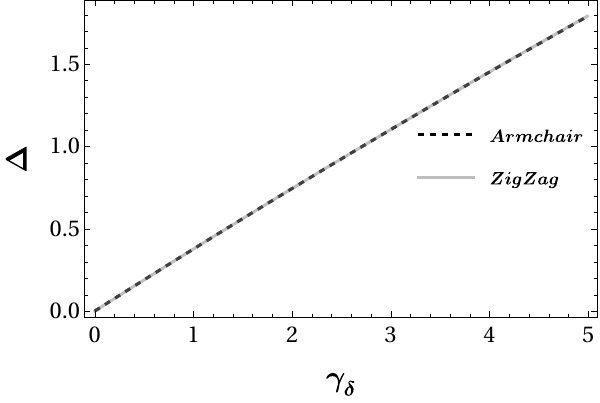}
  \caption{Band gap $\Delta$ as a function of the coupling strength $\gamma_\delta$ for zig-zag graphene nanoribbons (solid gray line) and arm-chair graphene nanoribbons (dashed black line).}
		\label{Gapvslambda}
\end{figure}

Since each channel can be analyzed independently, we use the trace of its $2\times2$ transport unit cell transfer matrix to identify the allowed and forbidden energy regions. We define

\begin{equation}
	a_{R,ZZ/AC}^{(i)}(E) = \frac{\mathrm{Re}\left[\mathrm{Tr}\, M^{(i)}_{ZZ/AC}(E)\right]}{2}
	\label{eq:kramer_channel} \, ,
\end{equation}
where the superscript $i$ denotes the transport channel. According to Kramer's condition~\cite{pereyra2002}, the allowed energy regions satisfy $|a^{(i)}_{R,ZZ/AC}(E)|<1$, whereas $|a^{(i)}_{R,ZZ/AC}(E)|>1$ corresponds to forbidden energy regions. We extract $\Delta(\gamma^{}_\delta)$ for both the arm-chair and zig-zag configurations using Eq (\ref{eq:kramer_channel}). In Fig. \ref{Gapvslambda}, we show the gap $\Delta$ as a function
of the scatterer strength $\gamma^{}_\delta$ for zig-zag and
arm-chair graphene nanoribbons
\begin{equation}\label{Gap}
	\Delta = 0.364265 \gamma^{}_\delta \hspace{0.3in} \gamma^{}_\delta >0 ,
\end{equation}
which describes both configurations. Thus, within the present model, the low-energy gap-opening branch does not show a systematic dependence on the edge orientation, in agreement with the experimental observations reported by Han et al. \cite{han2007}.

\section{Conclusions}
\label{sec:conclusions}

In this work, we developed a transfer-matrix formalism based on the theory of finite periodic systems to study electronic transport and gap formation in graphene nanoribbons with arm-chair and zig-zag edges. Each carbon atom was modeled as a localized $\delta$-scatterer with interaction strength $\gamma_{\delta}$, so that the forbidden energy regions arise from coherent multiple scattering in the finite periodic structure.

The transmission spectra show the resonant behavior expected for finite periodic systems. The number of resonances inside each allowed band increases with the number of unit cells and follows the relation
$N_{\mathrm{res}} = 2 n - 1,$
in agreement with the theory of finite periodic systems. These oscillations are associated with constructive and destructive interference between waves scattered multiple times within the periodic structure.

For $\gamma_{\delta}=0$, the structure is transparent and there is no gap. For repulsive interactions, $\gamma_{\delta}>0$, a low-energy gap opens and increases linearly with the scatterer strength according to
$\Delta = 0.364265\,\gamma_{\delta}$.
The same relation is obtained for arm-chair and zig-zag nanoribbons, indicating that the low-energy transport gap is independent of the crystallographic edge orientation within the present model. Our result that arm-chair and zig-zag nanoribbons exhibit the same low-energy transport gap is qualitatively consistent with the experiments of Han et al. \cite{han2007}, where the measured gap was found to depend primarily on ribbon width, with no clear dependence on crystallographic orientation. However, our analysis shows that the full transmission spectrum remains sensitive to edge geometry, as arm-chair and zig-zag nanoribbons exhibit different resonance amplitudes and distinct forbidden-energy intervals at higher energies.

The analytical results presented here reveal a fundamental difference between the two edge configurations. For arm-chair nanoribbons, the off-diagonal coupling coefficients vanish exactly, so that the transport channels remain completely decoupled. In contrast, for zig-zag nanoribbons, the interchannel coupling is finite in narrow ribbons but decreases with increasing ribbon width, approaching zero in the wide-ribbon limit. This behavior originates from the distinct atomic arrangements at the ribbon edges, demonstrating that the edge termination plays a crucial role in determining the coupling between transverse modes. Although the low-energy gap obtained here is identical for both edge configurations, the different interchannel couplings may influence the detailed transmission spectrum. Moreover, structural imperfections, edge disorder, and lattice impurities can enhance the mixing between transverse modes and may eventually modify the effective gap, as has been reported in the literature \cite{mucciolo2009,raygoza2022}. Notably, since the exact vanishing of the off-diagonal coupling in arm-chair nanoribbons relies on perfect positional periodicity and a spatially uniform scatterer strength, disorder can break this protection as well. Consequently, disorder-induced interchannel mixing is not restricted to zig-zag edges, in contrast to the intrinsic, width-dependent coupling discussed above.

Therefore, the $\delta$-scatterer model provides a simple and consistent description of gap opening and resonant transport in graphene nanoribbons. The parameter $\gamma_{\delta}$ represents an effective local scattering strength and may be determined, within the present model, by comparison with experimental gap measurements.

\begin{acknowledgments}
This work was supported by DCB UAM-A grant numbers
22322035, and 22322036. PPP and JCSS acknowledge the financial support of SECIHTI, SNII under CVU Nos. 5329 and 337363. CIAJ acknowledges the financial support of SECIHTI through a Master’s scholarship (CVU No. 993928).
\end{acknowledgments}

\appendix
\section{Derivation of the coupling coefficients}
In this appendix, we present the detailed derivation of the coupling coefficients
$\Gamma_{l,ij}$ for the zig-zag and arm-chair configurations. We begin with
\begin{equation}
	\Gamma_{l,ij}=\frac{4m^*_{e}\gamma_\delta}{\hbar^2 W_{y}}\sum_{\mu}
	\sin{\left(\frac{n_{i}\pi}{W_{y}}y_{\mu,l}\right)}
	\sin{\left(\frac{n_{j}\pi}{W_{y}}y_{\mu,l}\right)}.
\end{equation}
Here $l$ labels the lines of scatterers within the transport unit cell ($l=1,2$; the arm-chair cell repeats each line twice, Eq.~(\ref{eq:MAC})). To illustrate the calculation, we carry out the derivation explicitly for the
arm-chair nanoribbon; the zig-zag case follows analogously. We first evaluate the
sum over the atomic sites for the odd atomic lines,
\begin{equation}
	\Lambda_{l,ij}=\sum_{\mu=0}^{N_y-1}
	\sin{\left(\frac{n_{i}\pi}{W_{y}}y_{\mu,l}\right)}
	\sin{\left(\frac{n_{j}\pi}{W_{y}}y_{\mu,l}\right)},
\end{equation}
using the trigonometric identity $\sin a \sin b= \frac{1}{2}[\cos(a-b)- \cos(a+b)]$
and the atomic positions given by Eq.~(\ref{posAC}),
\begin{equation}
\begin{aligned}
\Lambda_{l,ij}
&= \frac{1}{2}\sum_{\mu=0}^{N_y-1}
   \Biggl[
   \cos\left(\frac{2\mu+1}{2N_y}\pi m_-\right)   - \cos\left(\frac{2\mu+1}{2N_y}\pi m_+\right)
   \Biggr]
\\[6pt]
&= \frac{1}{2}\,\mathrm{Re}\Biggl[
   e^{\frac{im_-\pi}{2N_y}}
   \sum_{\mu=0}^{N_y-1} r_-^\mu
   - e^{\frac{im_+\pi}{2N_y}}
   \sum_{\mu=0}^{N_y-1} r_+^\mu
   \Biggr].
\end{aligned}
\end{equation}
where $m_\pm = n_i \pm n_j$ and $r_\pm = \exp(im_\pm\pi/N_y)$. The geometric-series formula below requires $r_\pm \neq 1$, i.e.\ $m_\pm \neq 2kN_y$ for any integer $k$. For $n_i,n_j<N_y$ this holds for $m_+$, and for $m_-$ whenever $i\neq j$; the diagonal case $i=j$ ($m_-=0$, $r_-=1$) is treated separately, since the corresponding sum equals $N_y$ and the $m_+$ term vanishes. The condition $n_i,n_j<N_y$ means that the number of nodes of the transverse mode is smaller than the number of atoms per line along the $y$-direction. Under this condition, each sum
reduces to a finite geometric series, and the expression becomes
\begin{eqnarray}
	\Lambda_{l,ij}=\frac{1}{2}\,\mathrm{Re} \left[
	e^{\frac{im_-\pi}{2N_y}} \frac{1-r_-^{N_y}}{1-r_-}
	- e^{\frac{im_+\pi}{2N_y}} \frac{1-r_+^{N_y}}{1-r_+} \right],
\end{eqnarray}
which evaluates to
\begin{equation}
	\Lambda_{l,ij} = \frac{N_y}{2}\,\delta_{ij}.
\end{equation}
The same procedure applied to the even scatterer line yields an identical result.

For the zig-zag configuration, an analogous calculation is
performed. The sum now runs over $\mu = 0, 1, \ldots, 2N_y - 1$
atomic sites, with transverse positions $y_{\mu,l}$ given by
Eq.~(\ref{posZZ}). Applying the same product-to-sum identity as
in the arm-chair case, the channel-coupling sum is
\begin{align}
	\Lambda_{1,ij} = \frac{1}{2}
	\sum_{\mu=0}^{2N_y - 1}
	\left[
	\cos\!\left(\frac{m_-\,\pi\, y_{\mu,l}}{W_y}\right)
	- \cos\!\left(\frac{m_+\,\pi\, y_{\mu,l}}{W_y}\right)
	\right].
	\label{eq:LambdaZZ}
\end{align}

Due to the non-uniform spacing between nearest-neighbor carbon
atoms in the zig-zag geometry, the atomic positions $y_{\mu,l}$
are not evenly distributed. Consequently, the sum in
Eq.~(\ref{eq:LambdaZZ}) cannot be cast directly as a single
geometric series. Instead, we split the summation according to
the parity of $\mu$, separating the even sites ($\mu = 2k$)
from the odd sites ($\mu = 2k+1$), yielding
\begin{equation}
\label{eq:CmAB}
\begin{aligned}
C(m_\pm)
&= \underbrace{
   \sum_{k=0}^{N_y-1}
   \cos\!\left(
   \frac{m_\pm\pi(6k+3)}{6N_y}
   \right)
   }_{C_1\;(\mu\ \text{odd})}
\\[10pt]
&\quad + \underbrace{
   \sum_{k=0}^{N_y-1}
   \cos\!\left(
   \frac{m_\pm\pi(6k+1)}{6N_y}
   \right)
   }_{C_2\;(\mu\ \text{even})}.
\end{aligned}
\end{equation}

\noindent so that $\Lambda_{1,ij} = C(m_-) - C(m_+)$. The
explicit arguments $(6k+1)$ and $(6k+3)$ reflect the two
inequivalent atomic spacings within each unit cell of the
zig-zag edge. Each of the sublattice sums $C_1$ and $C_2$
is then evaluated as a geometric series following the same
procedure used for the arm-chair configuration. The final result
depends on the parity of $m_\pm$. For $i = j$,

\begin{equation}
	\Lambda_{l,ii} = N_y,
\end{equation}

\noindent while for $i \neq j$ with both $m_-$ and $m_+$ even,

\begin{equation}
	\Lambda_{l,ij} = 0,
\end{equation}

\noindent and for $i \neq j$ with both $m_-$ and $m_+$ odd,

\begin{equation}
	\Lambda_{l,ij} = \frac{1}{2}\left[\frac{\sin\dfrac{m_-\pi}{3N_y}}%
	{\sin\dfrac{m_-\pi}{2N_y}}
	-\frac{\sin\dfrac{m_+\pi}{3N_y}}%
	{\sin\dfrac{m_+\pi}{2N_y}}\right].
	\label{eq:LambdaResult}
\end{equation}

This result holds for both scatterer lines in the
zig-zag unit cell.

\begin{figure}[ht]
	\begin{center}
		\includegraphics[angle=0,width=\linewidth]{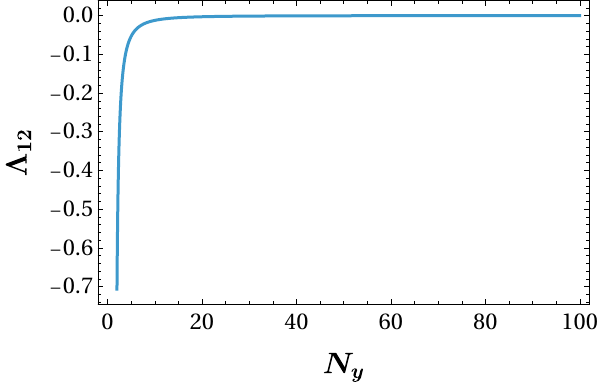}
		\caption{$\Lambda_{12}$ for case of two channels for zig-zag configuration as a function of $N_y$ for $i \neq j$ with $m_-=1$ and $m_+=3$ }
		\label{LambdaijvsNy}
	\end{center}
\end{figure}

We note that $\Lambda_{l,ii}\propto N_y$ for both edge
configurations. For arm-chair nanoribbons, the off-diagonal
elements vanish exactly, so that $\Lambda_{l,ij}=0$ for $i\neq j$.
For zig-zag nanoribbons, the off-diagonal elements approach zero
for sufficiently large $N_y$, as shown in
Fig.~\ref{LambdaijvsNy} for the two-channel case.

Consequently, for arm-chair nanoribbons, $\Gamma_{l,12}=\Gamma_{l,21}=0$, whereas for sufficiently wide zig-zag nanoribbons, $\Gamma_{l,12}\approx\Gamma_{l,21}\approx0$. Under these conditions the channels decouple. Reordering the amplitudes by channel, $(A_1,A_2,B_1,B_2)\to(A_1,B_1,A_2,B_2)$, a similarity transformation that leaves traces and the criterion of Eq.~(\ref{eq:kramer_channel}) unchanged, $M_{\delta,l}$ and $m_1$ become block diagonal,
\begin{equation}
	M_{\delta,l}=\begin{pmatrix}
		1-i\displaystyle{\frac{\Gamma_{l,11}}{2k_1}}& -i\displaystyle{\frac{\Gamma_{l,11}}{2k_{1}}} & 0 & 0 \\
		i\displaystyle{\frac{\Gamma_{l,11}}{2k_{1}}} & 1+i\displaystyle{\frac{\Gamma_{l,11}}{2k_{1}}} & 0 & 0\\
		0 & 0 & 1-i\displaystyle{\frac{\Gamma_{l,22}}{2k_{2}}} & -i\displaystyle{\frac{\Gamma_{l,22}}{2k_{2}}}\\
		0 & 0 &i\displaystyle{\frac{\Gamma_{l,22}}{2k_{2}}} & 1+i\displaystyle{\frac{\Gamma_{l,22}}{2k_{2}}}\\
	\end{pmatrix} \, ,
\end{equation}
\begin{eqnarray}
	m_{1}&=&\begin{pmatrix}
		e^{ik_{1} d/2}& 0 & 0 & 0 \\
		0 & e^{-ik_{1} d/2} & 0 & 0 \\
		0 & 0 & e^{ik_{2} d/2} & 0\\
		0 & 0 & 0 & e^{-ik_{2}d/2}\\
	\end{pmatrix} \, .
\end{eqnarray}
where each $2\times2$ block describes one channel.

\bibliographystyle{elsarticle-num}
\bibliography{biblio.bib}
\end{document}